\documentclass[12pt]{article}

\title{Algebraic Nonlinear Collective Motion}
\author{ J. Troupe and G. Rosensteel  \\  
Physics Department, Tulane University, New Orleans, LA 70118}
\date{Revision: July 15, 1998}

\begin{document}
\maketitle
\begin{abstract}
Finite-dimensional Lie algebras of vector fields determine geometrical collective models in quantum and classical physics. Every set of vector fields on Euclidean space that generates the Lie algebra sl(3,{{\bf R}}) and contains the angular momentum algebra so(3) is determined. The subset of divergence-free sl(3,{{\bf R}}) vector fields is proven to be indexed by a real number $\Lambda$. The $\Lambda=0$ solution is the linear representation that corresponds to the Riemann ellipsoidal model. The nonlinear group action on Euclidean space transforms a certain family of deformed droplets among themselves. For positive $\Lambda$, the droplets have a neck that becomes more pronounced as $\Lambda$ increases; for negative $\Lambda$, the droplets contain a spherical bubble of radius $|\Lambda|^{\frac{1}{3}}$. The nonlinear vector field algebra is extended to the nonlinear general collective motion algebra gcm(3) which includes the inertia tensor. The quantum algebraic models of nonlinear nuclear collective motion are given by irreducible unitary representations of the nonlinear gcm(3) Lie algebra. These representations model fissioning isotopes ($\Lambda>0$) and bubble and two-fluid nuclei ($\Lambda<0$).
\end{abstract}

\newpage

\section{Introduction}
The geometrical models of nuclear rotations and vibrations are based upon the general linear group GL(3,{\bf R}), the group of invertible linear transformations of three dimensional Euclidean space, and its subgroups, the special linear group SL(3,{\bf R}) and the rotation group SO(3) \cite{Tomonaga55,Cusson68,Weaver72}. These three groups are the motion groups of their corresponding models, the general collective motion model GCM(3) \cite{Rose79,Rose88}, the special collective motion model SCM(3) \cite{Rose76,Weaver76}, and the adiabatic rotational model ROT(3) \cite{Ui70,Weaver73}.

A collective model describes a dynamical system if the particle trajectories are compatible with the model's motion group. For example, the quantum adiabatic rotational model and the classical Euler rigid body model are useful when the velocity $\vec{v}$ of a particle located at the position vector $\vec{x}$ is given by $\vec{v} = \vec{\omega}\times \vec{x}$, where $\vec{\omega}$ is the angular velocity of the rigidly rotating system. In this case we say that the particle dynamics is compatible with the rotation group SO(3) since the velocity field is constrained to be an element of the Lie algebra so(3) of the rotation group, i.e., $v_{i} = \Omega_{ij} x_{j}$, where the $3\times 3$ matrix $\Omega_{ij} = \epsilon_{ijk}\omega_{k}$ is an antisymmetric matrix. Recall that the set of all $3\times 3$ real antisymmetric matrices forms the Lie algebra of the rotation group.

The motion group for the quantum Bohr-Mottelson collective model \cite{Bohr52} and the classical Riemann ellipsoidal model \cite{Chandrasekhar69} is the general linear group GL(3,{\bf R}). The general linear group constraint means that the velocity field of the particle system is a linear function of the Cartesian position coordinates, i.e., $v_{i} = \Xi_{ij} x_{j}$, where $\Xi$ is a constant $3\times 3$ real matrix. Thus $\Xi$ is an element of the Lie algebra gl(3,{\bf R}), the set of all $3\times 3$ real matrices. Rigid rotation requires that $\Xi = \Omega$ be antisymmetric, while irrotational flow ($\nabla \times \vec{v} = 0$) demands that $\Xi$ be symmetric. Divergence-free vector fields ($\nabla \cdot \vec{v} = 0$), corresponding to incompressible flow, imply that $\Xi$ is traceless, i.e., $\Xi$ is an element of the Lie algebra sl(3,{\bf R}) of the special linear group.

A serious limitation imposed on these familiar geometrical models is that the velocity fields are always {\it linear}. This strong condition is violated unambiguously in some circumstances, e.g., fissioning isotopes with a neck or other exotic shapes, and possibly broken strongly for other nuclear states, even at low energy. The article has two interrelated objectives: first, to determine nonlinear vector fields on Euclidean space that, for example, may model the dynamics of a fissioning isotope, and, second, to construct collective models compatible with such nonlinear motion groups.

The first problem is a surfeit of riches: there is an $\aleph$-infinity of nonlinear vector fields from which to choose the model's collective dynamics. Indeed every triple of smooth functions on Euclidean space defines a different vector field. Among all possible vector fields, a subset $\cal V$ of collective motion vector fields will be chosen. To be relevant to a dynamical system of interacting particles, the subset must assimilate the tangents to the particle trajectories; to be useful, these collective vector fields should not include much more.

The collective motion vector fields satisfy three general properties. First, $\cal V$ contains the angular momentum algebra so(3). Because of the SO(3) invariance of the nuclear Hamiltonian, rotated states are degenerate in energy. Thus collective rotational motion is expected as a rule. An exception to this typical situation occurs when rotational symmetry is broken, say by an external magnetic field. In this special case the set of collective vector fields $\cal V$ may contain only so(2).

Second, the collective velocity fields must form a finite-dimensional set. By definition, ``collective" particle motion means that the particle velocity vectors are constrained. Indeed collective motion requires that the particles move together in an orchestrated manner, and, therefore, the physically admissible particle velocity vectors are restricted to a finite-dimensional set. If the set $\cal V$ were infinite dimensional, then no collective physics would be infused into the model's ansatz.  At best, the complexity of the model would be similar or identical to that of the original many-body problem. As a practical matter, the dimension of the set of collective vector fields should be much less than the dimension of the A-particle Euclidean configuration space, $\dim {\cal V} << 3A$.

Finally, $\cal V$ is a Lie algebra, i.e., the set of collective vector fields is closed under commutation. One reason is that the family of collective motions is closed under composition. If $X$ is a collective vector field, let $\exp (t X)\vec{x}$ denote its integral curve through the point $\vec{x}$. If $X$ and $Y$ are two collective vector fields, then an admissible collective motion through $\vec{x}$ is given by the composition of collective modes
\begin{equation}
\gamma(t) = \exp(\sqrt{t}X)  \exp(\sqrt{t}Y)  \exp(-\sqrt{t}X)  \exp(-\sqrt{t}Y) \vec{x} . 
\end{equation}
Since the tangent to the curve $\gamma$ is the vector field $[X,Y]$, the commutator of two collective vector fields must also be a collective vector field.

 Another reason is that this algebraic property is necessary to formulate a quantum model of collective motion corresponding to the geometric vector fields in $\cal V$. The natural construction of the quantum theory from the Lie algebra of collective vector fields is given by a series of well-defined mathematical operations; the archetypes for the construction are the microscopic theories corresponding to the adiabatic rotational and Bohr-Mottelson models \cite{Villars70, Zickendraht71, Dzublik72, Rowe80, Buck79}. First, the vector fields on Euclidean space lift to vector fields on the $A$-particle configuration space {\bf R}$^{3A}$ by simply summing over the particle index to create one-body operators. We seldom distinguish between $\cal V$, considered as vector fields on {\bf R}$^{3}$, and the isomorphic algebra of vector fields on {\bf R}$^{3A}$. The sum over all particles insures that both the Pauli principle is obeyed and that the generated motion is collective. 

Second, according to Frobenius's theorem, the collective vector fields on {\bf R}$^{3A}$ may be integrated to surfaces in the $A$-particle configuration space \cite{Warner83}. The vector fields in $\cal V$ are tangent everywhere to these integrated surfaces. Each surface is an orbit of a connected Lie group $G$ whose Lie algebra is $\cal V$. Third, each orbit is a homogeneous space diffeomorphic to the coset space $G/H$, where $H$ is the isotropy subgroup of the orbit surface. Excluding a set of measure zero, the isotropy subgroup is the same for all orbits; the exceptions are orbits of smaller dimension than the generic orbit. The common dimension of the generic orbit surface is the difference between the dimensions of $G$ and its subgroup $H$, $\dim G/H = \dim G - \dim H$. If the dimension of the $A$-particle configuration space is greater than the dimension of $\cal V$, $3A > \dim \cal V$, then the generic isotropy subgroup $H$ is expected typically to be discrete.

Fourth, a transversal submanifold $\cal N$ of {\bf R}$^{3A}$ is chosen that intersects each generic orbit orthogonally exactly once. The transversal manifold is not unique. After removing a set of measure zero (the lower dimensional G-orbits), the $A$-particle configuration space is diffeomorphic to the Cartesian product of $G/H$ and the transversal manifold $\cal N$. Since sets of measure zero are irrelevant to the Hilbert space of square-integrable measurable functions on {\bf R}$^{3A}$, this Hilbert space is isomorphic to the tensor product,
\begin{equation}
 {\cal L}^{2}({\bf R}^{3A}) = {\cal L}^{2}(G/H) \otimes {\cal L}^{2}(\cal N) .
\end{equation}
The collective vector fields of $\cal V$ act only on the first component ${\cal L}^{2}(G/H)$ of the tensor product; this space is the Hilbert space of collective wave functions. Intrinsic wave functions are vectors in the space ${\cal L}^{2}(\cal N)$. Bands of collective states are given by monomials $\phi_{\tau K I M} \otimes \chi$, where $\chi$ is a fixed intrinsic state in the ${\cal L}^2$ space of the transversal submanifold and $\phi_{\tau K I M}$ are bands of collective states from the ${\cal L}^2$ space of the homogeneous manifold $G/H$.

A familiar example illustrates this somewhat abstract construction: let $\cal V$ be the angular momentum algebra so(3) and suppose there is only one particle $A=1$. In this case the collective vector fields integrate to the surfaces of spheres $S_2$; the angular momentum vectors are tangent to the spheres. Each sphere is an orbit of the rotation group SO(3) and the sphere is diffeomorphic to the coset space $S_2 \cong SO(3)/SO(2)$. An exception is the singular orbit consisting of a single point, the origin. An orthogonal transversal is any straight line through the origin, e.g., the positive $z$-axis. Thus the Hilbert space of square-integrable wave functions is isomorphic to the tensor product
\begin{equation}
 {\cal L}^{2}({\bf R}^3) = {\cal L}^{2}(SO(3)/SO(2)) \otimes {\cal L}^{2}({\bf R}^+) .
\end{equation}
A basis of ``collective" wave functions consists of the spherical harmonics; intrinsic states are just radial wave functions.

For a successful decomposition into collective and intrinsic wave functions, it is absolutely essential that Frobenius's theorem may be applied. But the necessary and sufficient condition for the existence of an integrating surface is that the set of collective vector fields close under commutation. Hence there is no flexibility here: $\cal V$ must be a Lie algebra.

The mathematical problem of classifying all nonlinear vector fields on Euclidean space that close to form a finite dimensional Lie algebra was studied first by Sophus Lie \cite{Lie1880}. In one dimension there are only three possibilities: dilations ($ax$), affine maps ($ax+b$), and linear fractional transformations ($(ax+b)/(cx+d)$). In two dimensions there is already a large number of possible nonisomorphic Lie algebras of vector fields. In three dimensions, because the number is so large, there is no known enumeration of all solutions. Lie's classification, however, is at once too fine and too coarse. It is too fine because we restrict to Lie algebras of vector fields that contain the angular momentum. It is too coarse because Lie does not distinguish among vector fields that differ by changes of variables: these are isomorphic Lie algebras of vector fields.  But physics does distinguish among velocity vectors that differ by point transformations since the particle trajectories are different. Indeed physics distinguishes among Hamiltonians that differ by canonical point transformations. Thus Lie's classic work is not directly relevant to our physics problem.

As an initial investigation of nonlinear collective nuclear motion, the smallest simple Lie algebra of smooth vector fields on Euclidean space that contains the angular momentum algebra will be determined. This minimal extension of so(3) is isomorphic to the algebra of the special linear group, ${\cal V} \simeq sl(3,{\bf R}) \supset so(3)$. It will be shown in section two that the set of all sl(3,{\bf R}) nonlinear vector fields is indexed by the set of smooth real-valued functions in one real variable. If each nonlinear vector field is also divergence free (because the motion is incompressible), then the set of incompressible nonlinear sl(3,{\bf R}) vector fields is indexed by a single real number $\Lambda$.

Among the solutions is the usual linear representation of sl(3,{\bf R}) corresponding to $\Lambda = 0$. The linear vector field is defined by the Lie derivatives of the natural linear action of the group SL(3,{\bf R}) on three dimensional Euclidean space. In section three it is proven that all incompressible nonlinear sl(3,{\bf R}) vector fields are given by certain nonlinear point transformations applied to the linear representation. The spaces of square integrable functions in the nonlinear geometries ($\Lambda \neq 0$) are shown to be isometric to the usual Hilbert space of square integrable wave functions ($\Lambda = 0$).

In section four the algebra of nonlinear sl(3,{\bf R}) vector fields is extended to the Lie algebra of gl(3,{\bf R}). A family of surfaces is discovered that is invariant with respect to the nonlinear GL(3,{\bf R}) motion group. These surfaces play the same role in the physical interpretation of the nonlinear theory as the inertia ellipsoids do in the linear case. The nonlinear surfaces include shapes with a neck ($\Lambda > 0$) and also configurations with a bubble ($\Lambda < 0$).

In section five the construction of nonlinear collective models is achieved. A closed fifteen dimensional Lie algebra, isomorphic to gcm(3), is presented that contains both the gl(3,{\bf R}) algebra of the nonlinear motion group plus certain nonlinear quadrupole and monopole tensors. When $\Lambda = 0$ the nonlinear gcm(3) algebra simplifies to the usual linear gcm(3) algebra of the Riemann ellipsoidal model. All nonlinear collective models compatible with the nonlinear GL(3,{\bf R}) motion group are given by the irreducible unitary representations of the GCM(3) Lie group. These representations are determined completely by the inducing construction because GCM(3) is a semidirect product with an abelian normal subgroup.

Because of the relevance to the nonlinear theory of the natural linear representation of the general linear group on three dimensional Euclidean space, a synopsis of its properties is provided here to close the introduction. The general linear group  contains the familiar rotation group SO(3) as a subgroup. In addition, GL(3,{\bf R}) contains the subgroup of invertible diagonal matrices that transform space by expanding and compressing the Euclidean $x-y-z$ axes lengths. Yet another subgroup is the nilpotent group of upper triangular matrices with ones running down the diagonal; these group elements are the shear transformations. The special linear group SL(3,{\bf R}) of unimodular matrices is a subgroup of the general linear group; SL(3,{\bf R}) consists of the incompressible linear transformations \cite{Cusson68,Weaver72}. An important property of the general linear group is that its action transforms an ellipsoid into another (deformed and rotated) ellipsoid. This property is at the foundation of the Riemann ellipsoidal model \cite{Rose88,Chandrasekhar69}.  

The unitary representation $U$ of the general linear group on the Hilbert space of A-nucleon wave functions is provided by
\begin{equation}
(U(g)\Psi)(\vec{r}_1,\vec{r}_2,\ldots,\vec{r}_A) = (\det g)^{1/2} \Psi(g^{-1}\cdot\vec{r}_1,\, g^{-1}\cdot\vec{r}_2,\, \ldots,g^{-1}\cdot\vec{r}_A) ,
\end{equation}
where $g\in$ GL(3,{\bf R}). The quantum mechanical hermitian observable that corresponds to each one-parameter geometrical group action on Euclidean space is defined by the Lie derivative of the representation $U$. The angular momentum operators, $L_{ij} = x_i p_j - x_j p_i$, summed over the particle index, span the Lie algebra of the rotation group. The breathing mode oscillation operator $S$ is the Lie derivative of the GL(1,{\bf R}) unitary representation of uniform scaling transformations
\begin{equation}
S =  \vec{r}\cdot\vec{p} - (3/2)\imath \hbar, 
\end{equation}
summed over the A-particles. For the transformation $g = \mbox{\rm diag}(e^\epsilon, e^\epsilon, e^\epsilon)$ of Euclidean space, the corresponding unitary operator is given by $U(g) = \exp (-\imath \epsilon S/\hbar)$.

The SL(3,{\bf R}) subgroup generates the angular momentum and a second-rank tensor operator $T$,
\begin{equation}
T_{ij} = x_i p_j + x_j p_i - (2/3) \vec{r}\cdot \vec{p} \, \delta_{ij} , \label{linearcart}
\end{equation}
where $1\leq i,j \leq 3$ label the Cartesian axes $x,y,z$ and the expression must be summed over the A-particles. To attain a geometrical understanding, note that the unimodular transformation 
\begin{equation}
g = \mbox{\rm diag}(e^{-\epsilon}, e^{-\epsilon}, e^{2\epsilon}) \label{subgp1}
\end{equation}
corresponds to the unitary operator $U(g) = \exp (-3 i \epsilon T_{33}/2 \hbar)$. There are similar expressions for incompressible scaling in the $x$ and $y$ directions. The off-diagonal components of the symmetric tensor $T_{ij}$ are associated with the hyperbola mappings
\begin{equation}
g = \left( \begin{array}{ccc}
   \cosh \epsilon & \sinh \epsilon & 0 \\
   \sinh \epsilon & \cosh \epsilon & 0 \\
   0 & 0 & 1
   \end{array} \right) \label{subgp2}
\end{equation}
for which $U(g) = \exp (-i \epsilon T_{12}/\hbar)$.

\section{Nonlinear Vector Fields on {\bf R}$^3$}
In this section all smooth vector fields on three dimensional Euclidean space are determined that (1) obey the sl(3,{\bf R}) commutation relations and (2) contain the standard angular momentum so(3) subalgebra. The results are summarized in the theorem at the end of this section. The linear vector field representation of sl(3,{\bf R}), as defined by the $T_{ij}$'s above, is one acceptable solution. Is it the unique solution to conditions (1) and (2), or are there other nonlinear solutions?

Were it not for the so(3) subalgebra limitation, the exhaustive computation of all sl(3,{\bf R}) vector fields would be a formidable task with many solutions. When the so(3) constraint is imposed, the set of all smooth nonlinear sl(3,{\bf R}) vector fields will be proven to be indexed by a single smooth function $A(r)$ of the radial coordinate $r$. The linear solution is just one of the possibilities, $A(r)=r$. If the sl(3,{\bf R}) vector fields are restricted further to have zero divergence, then it will be shown that $A(r) = r + \Lambda / r^2$ where $\Lambda$ is a real constant.

Because the so(3) subalgebra is fixed, spherical coordinates and bases are most convenient for the calculations. The eight dimensional simple Lie algebra sl(3,{\bf R}) is spanned by the vector angular momentum $L$ and a second rank tensor $T$. The angular momenta span so(3), the maximal compact subalgebra of the unimodular algebra. The angular momentum vector fields are given in spherical coordinates $(r,\theta,\phi)$ by
\begin{eqnarray}
L_{0} & = & -i\frac{\partial}{\partial \phi} \nonumber \\
L_{\pm} & = & e^{\pm i\phi}\lbrack\pm \frac{\partial}{\partial \theta} + i \cot\theta\frac{\partial}{\partial \phi}\rbrack .
\end{eqnarray}
These vector fields obey the so(3) commutation relations
\begin{eqnarray}
[ L_{0}, L_{\pm} ] & = & \pm \, L_{\pm} \nonumber \\ 
\left[ L_{+}, L_{-} \right] & = & 2\, L_{0} .
\end{eqnarray}

In the spherical basis, the commutation relations require that the components  $T_{\mu}$ form a second rank tensor operator,
\begin{eqnarray}
[L_{0},T_{\mu}] & = & \mu \, T_{\mu} \nonumber \\
\left[ L_{\pm},T_{\mu} \right] & = & \sqrt{6-\mu(\mu\pm1)} \, T_{\mu\pm 1} ,
\end{eqnarray}
for $\mu = -2,\ldots,+2$, and that the set closes to form the simple Lie algebra sl(3,{\bf R}),
\begin{equation}
[T_{\mu},T_{\nu}] = -2\sqrt{10}\,(2{\mu}2{\nu} | 1{\mu+\nu})\,L_{\mu+\nu}, \label{Tall}
\end{equation}
where $L_{\pm 1} = \mp (1/\sqrt{2}) L_{\pm}$. The spherical basis is normalized to the Cartesian basis by 
\begin{equation}
T_{\pm 2} = \frac{1}{2}(T_{11}-T_{22}) \pm i T_{12}.
\end{equation}
In particular, the spherical component $T_{+2}$ of the linear vector fields of Eq.\,(\ref{linearcart}), is given by
\begin{equation}
T_{+2} = -i e^{2i\phi}\lbrack r\sin^2\theta\frac{\partial}{\partial r}+\frac{1}{2}\sin2\theta\frac{\partial}{\partial\theta} +i \frac{\partial}{\partial\phi}\rbrack \label{linearp2} .
\end{equation}

To find all possible sl(3,{\bf R}) vector fields, we start with a completely general form for $T_{+2}$,
\begin{equation}
T_{+2} = -i \, \lbrack a(r,\theta,\phi)\frac{\partial}{\partial r}+b(r,\theta,\phi)\frac{\partial}{\partial\theta}+c(r,\theta,\phi)\frac{\partial}{\partial\phi} \rbrack,
\end{equation}
and then impose the commutation relations upon it. The coefficients $a,b,c$ of the vector field are assumed to be smooth functions of the spherical coordinates in an open subset of Euclidean space that is contained within the domain of the spherical coordinate chart, viz., $r>0$, $0<\theta<\pi$, $0\leq \phi \leq 2\pi$, or {\bf R}$^3$ minus the $z$-axis. In addition each coefficient is periodic in $\phi$ with period $2\pi$.

First, $[L_{0}, T_{+2}] = 2\, T_{+2}$ determines completely the $\phi$ dependence of the coefficients,  
\begin{equation}
T_{+2} = -i\, e^{2i \phi}\lbrack a(r,\theta)\frac{\partial}{\partial r}+b(r,\theta)\frac{\partial}{\partial\theta}+c(r,\theta)\frac{\partial}{\partial\phi}\rbrack .
\end{equation}
Second, the highest weight condition, $[L_{+}, T_{+2}] = 0$, requires that three independent equations be satisfied, viz., the vanishing coefficients of the partial derivatives in the $r, \theta$, and $\phi$ directions,
\begin{eqnarray}
\frac{\partial a}{\partial\theta}(r,\theta) - 2\cot\theta \, a(r,\theta) &=& 0 \label{condition1} \\
\frac{\partial b}{\partial\theta}(r,\theta) - 2\cot\theta \, b(r,\theta) - ic(r,\theta) &=& 0 \label{condition2}  \\
\frac{\partial c}{\partial\theta}(r,\theta) - \cot\theta \, c(r,\theta)+ i\csc^{2}\theta \, b(r,\theta) &=& 0. \label{condition3}
\end{eqnarray}
Eq.\,(\ref{condition1}) can be solved immediately to give
\begin{equation}
a(r,\theta)=A(r) \sin^{2}\theta.
\end{equation}
If $b(r,\theta)$ is expressed in terms of $c(r,\theta)$ using Eq.\,(\ref{condition3}) and substituted into Eq.\,(\ref{condition2}), then the resulting second order differential equation may be solved for $c(r,\theta)$. The general solutions are 
\begin{eqnarray}
b(r,\theta) & = & i\,\sin\theta\lbrack\alpha(r) - \beta(r)\cos\theta \rbrack \nonumber \\
c(r,\theta) & = & -\alpha(r)\cos\theta + \beta(r)
\end{eqnarray}
where $A(r)$, $\alpha(r)$, and $\beta(r)$ are arbitrary smooth functions of $r$. In addition, repeated application of the lowering operator $L_{-}$ to the highest weight determines the lowest weight component of the tensor $T$,
\begin{eqnarray}
T_{\pm2} & = & -i e^{\pm2i\phi}\lbrack A(r)\sin^{2}\theta\frac{\partial}{\partial r}\pm i\sin\theta( \alpha(r)\mp\beta(r)\cos\theta)\frac{\partial}{\partial\theta} \nonumber \\
& & \pm(\beta(r)\mp\alpha(r)\cos\theta)\frac{\partial}{\partial \phi}\rbrack.
\end{eqnarray}
Because of the Jacobi identity and the fact that $T$ is a tensor operator, it is sufficient for the one commutation relation,
\begin{equation}
[T_{+2}, T_{-2}] = -4\, L_{0} , \label{Tp2m2}
\end{equation}
to be satisfied to guarantee that all the commutation relations among the components of T, Eq.\,(\ref{Tall}), are correct. Eq.\,(\ref{Tp2m2}) yields three equations,
\begin{eqnarray}
0 & = & A(r)\alpha(r)  \label{eq1}\\
0 & = &  A(r)\frac{d\alpha}{dr} \label{eq2}  - 3 i \alpha(r)\beta(r)\\
4i & = & -4i\beta(r)^{2}+4i\alpha(r)^{2}\cos^2\theta+2\left(A(r)\frac{d\beta}{dr}-i\alpha(r)^{2}\right)\sin^{2}\theta \label{eq3} .
\end{eqnarray}
Since Eq.\,(\ref{eq3}) must be satisfied for all $0<\theta<\pi$, the coefficients of $\cos^2\theta$ and $\sin^2\theta$ are zero while the constant term on the right hand side must equal $4i$. The solution is $\alpha(r)=0$ and $\beta(r)^2 = -1$ or $\beta(r) = \pm i$. With these choices for $\alpha$ and $\beta$, Eqs.(\ref{eq1},\ref{eq2}) are satisfied too for any function $A(r)$.
If $\beta=+i$, then assembling our results, 
\begin{equation}
T_{\pm2}= -i e^{\pm2i\phi}\lbrack A(r)\sin^{2}\theta\frac{\partial}{\partial r}+\frac{1}{2}\sin2\theta\frac{\partial}{\partial\theta}\pm i\frac{\partial}{\partial\phi}\rbrack. \label{lastform}
\end{equation}
for some smooth function $A(r)$ of the radial coordinate. If $A(r)=r$, then the linear representation of sl(3,{\bf R}) is recovered.

From Eq.\,(\ref{Tp2m2}) it is evident that, if $T_{\mu}$ is a solution to the sl(3,{\bf R}) commutation relations, then a multiple $c\, T_{\mu}$ is also a solution if and only if $c^2=1$ or $c=\pm 1$. Note that $\beta=\pm i$ corresponds to $c=\pm 1$.  For simplicity, the ambiguous phase was chosen in Eq.\,(\ref{lastform}) to match the conventional phase of the linear vector fields, Eq.\,(\ref{linearp2}).

Thus the set of smooth sl(3,{\bf R}) vector fields on ${\bf R}^3$ that contain the usual so(3) subalgebra of angular momentum vector fields is indexed by a smooth function $A(r)$. We now see how strict the angular momentum constraint really is. All the angular dependence of the $T_{\mu}$'s has been determined and the only freedom left is in the radial component.

The radial component may be constrained further if the motion is incompressible. In this case, the sl(3,{\bf R}) vector fields must be divergence free. But the divergence for the highest and lowest weight components is
\begin{equation}
\mbox{div\ } T_{\pm 2} = -i\, e^{\pm 2 i \phi} \sin^2\theta \left[ \frac{1}{r^2}\frac{\partial}{\partial r} (r^2 A(r)) -3 \right] .
\end{equation}
The divergence vanishes if
\begin{equation}
A(r) = r +\frac{\Lambda}{r^{2}} ,
\end{equation}
where $\Lambda$ is a real number with the dimensions of volume.
Summarizing the results, we have proven the following:

\vspace{.2in}
\noindent{\bf Theorem}. \ For every smooth function $A(r)$ of the radial coordinate $r$, there is a representation of the simple Lie algebra sl(3,{\bf R}) by smooth vector fields in an open domain of ${\bf R}^3$ such that (1) the so(3) rotation subalgebra is represented by the standard angular momentum operators and (2) the components of the rank two spherical tensor are given by 
\begin{eqnarray}
T_{\pm2}(A) &=& -i e^{\pm2 i\phi}\lbrack A(r) \sin^{2}\theta \frac{\partial}{\partial r}+\frac{1}{2}\sin2\theta\frac{\partial}{\partial\theta}\pm i\frac{\partial}{\partial\phi} \rbrack \nonumber \\
T_{\pm1}(A) &=& -i e^{\pm i\phi}\lbrack \mp A(r)  \sin2\theta \frac{\partial}{\partial r}\mp \cos2\theta\frac{\partial}{\partial\theta}- i\cot\theta\frac{\partial}{\partial\phi}\rbrack  \nonumber \\
T_{0}(A) &=&  \frac{-i}{\sqrt{6}}\lbrack A(r) (1+3\cos2\theta)\frac{\partial}{\partial r}-3\sin2\theta\frac{\partial}{\partial\theta}\rbrack . \nonumber
\end{eqnarray}
If the representation is restricted to divergence free vector fields, then $A(r)=r + \Lambda/r^2$, where $\Lambda$ is a real number. Note that $\Lambda=0$ is the usual linear representation.

As a corollary, there is no similar representation of the compact Lie algebra su(3). The reason is that, since the complexifications of sl(3,{\bf R}) and su(3) are isomorphic, the theorem determines every solution to the su(3) algebra too, viz., $i\,T_{\mu}$. This su(3) tensor corresponds to a ``vector field'' with pure imaginary components.



\section{Origin of nonlinear sl(3,{\bf R}) vector fields}

In this section every divergence-free so(3)-preserving nonlinear representation of sl(3,{\bf R}) is shown to be given by a change of variables applied to the linear representation. The proof is achieved using the concept of a manifold.

\vspace{.1in}
\noindent {\bf Definition.}\ A set $M$ is a smooth manifold if: \cite{Hicks}

\begin{enumerate}
\item $M$ is a topological space with a family of open sets $M_{\alpha}$ that cover it, $\bigcup_{\alpha}M_{\alpha}=M$.
\item Associated with each such open set is a one-to-one bicontinuous mapping $\psi_{\alpha}$ from $M_{\alpha}$ onto an open set ${\cal O}_{\alpha}$ in {\bf R}$^n$. Each pair $(M_{\alpha},\psi_{\alpha})$ is called a chart and the set of all charts is called an atlas for the manifold $M$.
\item Given two different charts, $(M_{\alpha}, \psi_{\alpha})$ and $(M_{\beta}, \psi_{\beta})$, the map $\psi_{\beta}\circ\phi_{\alpha}^{-1}$ from the open subset $\psi_{\alpha}(M_{\alpha}\cap M_{\beta})$ of ${\bf R}^n$ to the open subset $\psi_{\beta}(M_{\alpha}\cap M_{\beta})$ of ${\bf R}^n$ is smooth, i.e., infinitely differentiable.
\end{enumerate}
\vspace{.1in}

\noindent It is important to maintain the distinction between a manifold and the charts that provide its coordinatization. For the Euclidean space $M={\bf R}^3$ this distinction is often ignored because the usual atlas, Cartesian coordinates, consists of only a single chart $(M_{xyz},\psi_{xyz})$, where $M_{xyz}={\cal O}_{xyz}={\bf R}^3$, and $\psi_{xyz}: m \mapsto (x,y,z)$ is the identity mapping for every point $m\in{\bf R}^3$. Another common chart for three dimensional space is given by the spherical coordinates. But spherical coordinates do not supply a global chart for ${\bf R}^3$ because the homeomorphism $\psi_{r\theta\phi}: m \mapsto (r,\theta,\phi)$ is not one-to-one everywhere. Along the $z$-axis, $\theta =0$ and $\theta=\pi$, all values of the longitudinal angle $\phi$ correspond to the same point of the manifold $M={\bf R}^3$. The situation is even worse at the origin, $r=0$, where all angular coordinates $\theta$ and $\phi$ correspond to the same point of the manifold. In addition, since the longitudinal angle range $[0,2\pi]$ is a closed set, at least two overlapping charts are required to cover {\bf R}$^3$ minus the $z$-axis. For example, one chart  $(M_{\phi\neq0},\psi_{r\theta\phi})$ excludes $\phi=0$, while the other  $(M_{\phi\neq\pi},\psi_{r\theta\phi})$ excludes $\phi=\pi$. Although the chart mapping $\psi_{r\theta\phi}$ is the same for both regions, the domains and images of these charts are different:
\begin{eqnarray}
{\cal O}_{\phi \neq 0} & = & \psi_{r\theta\phi} (M_{\phi \neq 0})  =  \{ (r,\theta,\phi) | r>0, 0<\theta<\pi, 0<\phi<2\pi \} \nonumber \\
{\cal O}_{\phi \neq \pi} & = & \psi_{r\theta\phi} (M_{\phi \neq \pi})  =  \{ (r,\theta,\phi) | r>0, 0<\theta<\pi, -\pi<\phi<\pi \} .
\end{eqnarray}
Evidently similar charts $(M_{\phi\neq\phi_0},\psi_{r\theta\phi})$ may be defined by excluding an arbitrary longitudinal angle $\phi_0$ from the range  ${\cal O}_{\phi \neq \phi_0}$. Let $(M_{r\theta\phi},\psi_{r\theta\phi})$ denote any such spherical coordinate chart mapping onto the open set $\psi_{r\theta\phi}(M_{r\theta\phi}) =  {\cal O}_{r\theta\phi}$. The third condition in the definition of the manifold is satisfied by the smooth function $\psi_{xyz}\circ\, \psi_{r\theta\phi}^{-1}: {\cal O}_{r\theta\phi} \mapsto \psi_{xyz}(M_{r\theta\phi})$, where $x=r\cos\phi\sin\theta$, $y=r\sin\phi\sin\theta$, $z=r\cos\theta$.

As will be proven shortly, the incompressible nonlinear vector fields are described naturally by two charts $(M_+,\psi_{+})$ and$(M_-,\psi_{-})$ corresponding to $\Lambda >0$ and $\Lambda<0$, respectively.  The open set $M_{+} = M_{r\theta\phi}$ is identical to the domain of a spherical coordinate chart, but the range ${\cal O}_{+} = \psi_{+}(M_{+})$ excludes a sphere of radius $R = \sqrt[3]{\Lambda}$:
\begin{eqnarray}
\psi_+ \circ \psi_{r\theta\phi}^{-1} : {\cal O}_{r\theta\phi} & \longrightarrow & {\cal O}_{+} \nonumber  \\
(r, \theta, \phi) & \longmapsto & (r', \theta', \phi') = (\sqrt[3]{r^3+\Lambda}, \theta, \phi) , \label{nonlinmapplus}
\end{eqnarray}
where ${\cal O}_{+} = {\cal O}_{r\theta\phi}  \cap \tilde{S}_{R}$ and $\tilde{S}_{R}$ denotes the complement of the closed sphere of radius $R$. If $\Lambda < 0$, the domain of the chart $M_{-} = M_{r\theta\phi}  \cap \tilde{S}_{R}$ is a region excluding the sphere of radius $R = \sqrt[3]{|\Lambda|}$, but now the range is ${\cal O}_{r\theta\phi} = \psi_{-}(M_{-})$ (see Figure 1): 
\begin{eqnarray}
\psi_- \circ \psi_{r\theta\phi}^{-1} : {\cal O}_{+} & \longrightarrow & {\cal O}_{r\theta\phi}   \nonumber  \\
(r, \theta, \phi) & \longmapsto & (r', \theta', \phi') = (\sqrt[3]{r^3+\Lambda}, \theta, \phi) . \label{nonlinmapminus}
\end{eqnarray}

The next step is to demonstrate that the nonlinear vector fields of \S2 can be constructed in a simple way via a change of chart applied to the linear vector field representation. Suppose $(M_{\alpha},\psi_{\alpha})$ is any chart of the manifold and ${\cal O}_{\alpha} = \psi_{\alpha}(M_{\alpha})$ is an open subset of {\bf R}$^3$. The Lie algebra sl(3,{\bf R}) is represented by linear vector fields on the open set ${\cal O}_{\alpha}$. In particular, if spherical coordinates $(r', \theta', \phi')$ are used for the open set ${\cal O}_{\alpha}$, then the highest weight component of the rank two sl(3,{\bf R}) vector field is given explicitly by
\begin{equation}
T_{+2}^{\prime} = -i e^{2i\phi^{\prime}}\lbrack r^{\prime}\sin^2\theta^{\prime}\frac{\partial}{\partial r^{\prime}}+\frac{1}{2}\sin2\theta^{\prime}\frac{\partial}{\partial\theta^{\prime}} +i \frac{\partial}{\partial\phi^{\prime}}\rbrack.
\end{equation}

A vector field defined on ${\cal O}_{\alpha}$ is mapped back to $M_{\alpha}$ in the Euclidean space by $\psi_{\alpha}^{-1}$ and then expressed in terms of the usual spherical coordinates by $\psi_{r\theta\phi}\circ\psi_{\alpha}^{-1}$. Hence, the Euclidean space vector field, corresponding to $T_{+2}^{\prime}$ on ${\cal O}_{+}$, is
\begin{eqnarray}
(\psi_{r\theta\phi}\circ\psi_{+}^{-1})_{\ast}\, T_{+2}^{\prime} & = & -i e^{2i\phi}\lbrack r^{\prime} \sin^2\theta \frac{dr}{dr^{\prime}}\frac{\partial}{\partial r}+\frac{1}{2}\sin2\theta\frac{\partial}{\partial\theta} +i \frac{\partial}{\partial\phi}\rbrack \nonumber \\              
 &=& -i e^{2i\phi}\lbrack  (r + \frac{\Lambda}{r^2}) \sin^2\theta \frac{\partial}{\partial r} + \frac{1}{2}\sin2\theta\frac{\partial}{\partial\theta} +i \frac{\partial}{\partial\phi}\rbrack \nonumber \\
&  = & T_{+2}(\Lambda) , 
\end{eqnarray}
where  $(\psi_{r\theta\phi}\circ\psi_{+}^{-1})_{\ast} $ denotes the Jacobian of the transformation $\psi_{r\theta\phi}\circ\psi_{\alpha}^{-1}$.
Indeed every component of the nonlinear sl(3,{\bf R}) representation for $\Lambda>0$ defined on the open set $M_+ = M_{r\theta\phi}$ is given by the coordinate change
\begin{equation}
T_{\mu}(\Lambda) = (\psi_{r\theta\phi}\circ\psi_{+}^{-1})_{\ast}\, T_{\mu}^{\prime} .
\end{equation}   
If $\Lambda<0$, a similar argument demonstrates that the nonlinear representation on the open subset $M_- = M_{r\theta\phi}  \cap \tilde{S}_{\sqrt[3]{|\Lambda|}}$ of {\bf R}$^3$ is given by the above equation when $\psi_{+}$ is replace by $\psi_{-}$. Thus, for $\Lambda<0$, the nonlinear representation is defined in a region of the Euclidean space exterior to the sphere of radius $\sqrt[3]{|\Lambda|}$.

Note that, for any $\Lambda$, the so(3) subalgebra is preserved by the mappings $(\theta'=\theta, \phi'=\phi)$:
\begin{equation}
L_{\mu} = (\psi_{r\theta\phi}\circ\psi_{\pm}^{-1})_{\ast} L_{\mu}^{\prime} .
\end{equation} 

It is now clear why the nonlinear vector fields are a representation of the sl(3,{\bf R}) algebra: The Jacobian mapping $(\psi_{r\theta\phi}\circ\psi_{\alpha}^{-1})_{\ast}$ is a Lie homomorphism of the algebra of linear vector fields defined on ${\cal O}_{\alpha}$ into the algebra of, in general nonlinear, vector fields on ${\cal O}_{r\theta\phi}$ \cite{Hicks}.

The nonlinear vector fields may not be complete. The linear vector fields are certainly complete on {\bf R}$^3$, i.e., they can be integrated to one-parameter subgroups of the Lie group SL(3,{\bf R}), e.g., Eqs.( \ref{subgp1}, \ref{subgp2} ). If a one-parameter subgroup is contained entirely within the region ${\cal O}_{\alpha}$, then the nonlinear vector field is complete because the subgroup is mapped entirely within $M_{\alpha}$ by $\psi_{\alpha}^{-1}$. Otherwise the vector field may not be complete. For $\Lambda < 0$, the nonlinear vector fields on $M_-$  are complete since ${\cal O}_{-}= {\cal O}_{r\theta\phi}$. Hence, SL(3,{\bf R}) acts on $M_{-}$ as a Lie transformation group. But for positive $\Lambda$ the region ${\cal O}_{+}$ excludes the sphere of radius $\sqrt[3]{\Lambda}$, and integral curves of the linear representation will typically run into this sphere. In this case, the integral curves of the nonlinear sl(3,{\bf R}) vector fields form one-parameter semigroups instead of one-parameter groups. 

Thus the connection between linear and nonlinear vector field representations of sl(3,{\bf R}) is established. To complete the picture, wave functions in different coordinate charts must be related to each other by reformulating the chart mappings as isometries between different Hilbert spaces.

If $M$ is a manifold and $\omega$ is a volume element on it, then define the Hilbert space ${\cal L}^{2}(M)$ to be the set of all measurable complex-valued square-integrable wave functions on $M$:
\begin{equation}
{\cal L}^{2}(M) = \left\{ f : M \rightarrow {\bf C}\, | \, f \mbox{\ is measurable and} \int_{M} |f|^2 \omega < \infty \right\} .
\end{equation}
A wave function $f$ is expressed in a chart $(M_{\alpha}, \psi_{\alpha})$ as a complex-valued function $f_{\alpha} = f \circ \psi_{\alpha}^{-1} : {\cal O}_{\alpha}\rightarrow$ {\bf C}. Given two overlapping charts $(M_{\alpha}, \psi_{\alpha})$ and $(M_{\beta}, \psi_{\beta})$, the coordinate expressions for the wave function $f$ are related by a change of variables, $f_{\beta} = f_{\alpha} \circ (\psi_{\alpha}\circ\phi_{\beta}^{-1})$. The expressions for the volume element in the two charts are related by the determinant $J_{\alpha\beta}$ of the Jacobian of the coordinate transformation $\psi_{\alpha}\circ\phi_{\beta}^{-1}, \omega_{\alpha} = J_{\alpha\beta} \omega_{\beta}$. If a single coordinate chart covers the entire manifold with the possible exception of a set of measure zero, then the Hilbert space ${\cal L}^{2}(M)$ is isomorphic to the space
\begin{equation}
{\cal L}^{2}({\cal O}_{\alpha}) = \left\{ f_{\alpha} : {\cal O}_{\alpha} \rightarrow {\bf C}\, | \, f_{\alpha} \mbox{\ is measurable and} \int_{{\cal O}_{\alpha}} |f_{\alpha}|^2 \omega_{\alpha} < \infty \right\} ,
\end{equation}
where $\omega_{\alpha}$ denotes the volume element $\omega$ as expressed in terms of the chart's coordinates.
More generally, wave functions and integration on the manifold $M$ are pieced together from different charts using a partition of unity \cite{Hicks}.

For example, the Hilbert space ${\cal L}^{2}({\bf R}^3)$ of square-integrable wave functions on the three dimensional Euclidean manifold is isomorphic to the space of complex-valued measurable wave functions of the Cartesian coordinates $(x,y,z)$ that are square-integrable relative to the volume element $dx\,dy\,dz$. Since the $z$-axis is a set of measure zero, the space of complex-valued measurable functions of the spherical coordinates $(r,\theta,\phi)$ that are square-integrable on the region ${\cal O}_{r\theta\phi}$ relative to $r^2\sin\theta dr\,d\theta\,d\phi$ is also isomorphic to ${\cal L}^{2}({\bf R}^3)$.

Since $M_{+} = M_{r\theta\phi}$, the space ${\cal L}^{2}({\cal O}_{+})$ is also isomorphic to ${\cal L}^{2}({\bf R}^3)$,
\begin{equation}
{\cal L}^{2}({\cal O}_{+}) = \left\{ f_{+} : {\cal O}_{+} \rightarrow {\bf C}\, | \, f_{+} \mbox{\ is measurable and} \int_{\cal O_{+}} |f_{+}|^{2} \omega_{+} < \infty \right\} .
\end{equation}
The volume element in the ${\cal O}_{+}$ chart coordinates $(r^{\prime},\theta^{\prime},\phi^{\prime})$ is computed from the Jacobian of the transformation from the spherical ${\cal O}_{r\theta\phi}$ chart coordinates 
\begin{eqnarray}
\omega_{+} & = & r^2 dr \sin\theta d\theta d\phi \nonumber \\
& = & r^2 \frac{dr}{dr^{\prime}} dr^{\prime} \sin\theta d\theta d\phi \nonumber \\
& = & (r^{\prime})^2\sin\theta^{\prime} dr^{\prime}  d\theta^{\prime}  d\phi^{\prime} .
\end{eqnarray}
Since $\omega_{+}$ is the same as the spherical coordinate volume element, the nonlinear vector fields $T_{\mu}(\Lambda)$ for $\Lambda>0$ form a hermitian rank two tensor operator.

Each Hilbert space ${\cal L}^{2}({\cal O}_{+})$ for positive $\Lambda$ is therefore canonically isomorphic to the Hilbert space ${\cal L}^{2}({\bf R}^3)\cong{\cal L}^{2}({\cal O}_{r\theta\phi})$. The explicit isometry is given by 
\begin{eqnarray}
U_{\Lambda}: {\cal L}^2({\cal O}_{r\theta\phi})& \to & {\cal L}^2({\cal O}_{+}) \nonumber \\
U_{\Lambda}f & = & f\circ\left(\psi_{r\theta\phi}\circ\psi_{+}^{-1}\right) \nonumber \\
(U_{\Lambda}f)(r^{\prime},\theta,\phi) & = & f(\sqrt[3]{r^{\prime 3}-\Lambda},\theta,\phi) \label{Ulambda} .
\end{eqnarray}
The isometries obey the additive property
\begin{equation}
U_{\Lambda_{1}}U_{\Lambda_{2}} = U_{\Lambda_{1}+\Lambda_{2}},
\end{equation}
and the inverse mapping is $U_{-\Lambda}: {\cal L}^2({\cal O}_{+}) \to  {\cal L}^2({\cal O}_{r\theta\phi})$.

The infinitesimal generator for $U_{\Lambda}=\exp(-i \Lambda G)$ is given by
\begin{eqnarray}
(G f)(r^{\prime},\theta,\phi) &=& i \frac{d}{d\Lambda}(U_{\Lambda}f)(r^{\prime},\theta,\phi)\mid_{\Lambda=0} \nonumber \\
& = & i \frac{d}{d\Lambda}f(\sqrt[3]{r^{\prime 3}-\Lambda},\theta,\phi)\mid_{\Lambda=0} \nonumber \\
& = & -i \frac{1}{3r^{\prime 2}}\frac{\partial f}{\partial r^{\prime}}(r^{\prime},\theta,\phi) .
\end{eqnarray}
$G$ is hermitian because $U_{\Lambda}$  is an isometry. In terms of the hermitian radial momentum operator, $p_r = -i r^{-1} (\partial/\partial r) r$, the generator is
\begin{equation}
G = \frac{1}{3r} p_{r} \frac{1}{r} .
\end{equation}
The nonlinear sl(3,{\bf R}) operators may be expressed as a sum of the linear representation plus the product of the rank-two spherical harmonics times the generator G,
\begin{equation}
T_{\mu}(\Lambda) =  T_{\mu}(0) + \sqrt{\frac{96\pi}{5}}\Lambda Y_{\mu} G .
\end{equation}

The situation for negative $\Lambda$ is more subtle because the chart $(M_{-},\psi_{-})$ omits the sphere of radius $R=|\Lambda|^{1/3}$ from the Euclidean manifold. In general, a second chart is required to cover the interior of the sphere. However, in the special case of bubble nuclei, the single-particle wave functions must vanish inside a sphere of radius $R$. The subspace of bubble nuclei wave functions is isomorphic to 
\begin{equation}
{\cal L}^{2}({\cal O}_{-}) = \left\{ f_{-} : {\cal O}_{-} \rightarrow {\bf C}\, | \, f_{-} \mbox{\ is measurable and} \int_{\cal O_{-}} |f_{-}|^{2} \omega_{-} < \infty \right\} .
\end{equation}
The volume element $\omega_{-}$ is again the same as the spherical coordinate volume element, and, hence, the nonlinear vector fields $T_{\mu}(\Lambda)$ for $\Lambda<0$ also form a hermitian rank two tensor operator.
The isometry $U_{\Lambda}$ for negative $\Lambda$ is given by Eq.\,(\ref{Ulambda}), but the range is now ${\cal L}^2({\cal O}_{-})$. 

The sl(3,{\bf R}) results may be extended easily to nonlinear gl(3,{\bf R}) vector fields. The hermitian monopole operator of gl(3,{\bf R}) is given in a coordinate chart $(M_{\pm},\psi_{\pm})$ by
\begin{equation}
S^{\prime} = -i r^{\prime} \frac{\partial}{\partial r^{\prime}} - \frac{3}{2} i .
\end{equation}
On the Euclidean manifold, the corresponding nonlinear vector field is
\begin{eqnarray}
S(\Lambda) & = &  (\psi_{r\theta\phi}\circ\psi_{\pm}^{-1})_{\ast} S^{\prime} \nonumber \\
& = & -i r^{\prime}  \frac{d r}{d r^{\prime}} \frac{\partial}{\partial r} - \frac{3}{2} i  \nonumber \\
& = & S(0) + 3 \Lambda G.
\end{eqnarray}

\section{Deformed Droplets}

Consider an ellipsoid in three-dimensional Euclidean space. Its boundary, defined uniquely by a positive-definite symmetric $3\times3$ real matrix $q$, consists of the points that satisfy the quadratic relation
\begin{equation}
^{t}\vec{r} \, q^{-1}\,  \vec{r} =1,
\end{equation}
where $\vec{r}$ denotes the Cartesian position vector. If $q$ is a diagonal matix with entries $a^2$, $b^2$, and $c^2$, then the ellipsoid's principal axes are aligned with the Cartesian axes and its semi-axis lengths are $a$, $b$, and $c$.

With respect to the usual action of the general linear group, $\vec{r}\rightarrow g\,\vec{r}$ for $g\in$\,GL(3,{\bf R}), an ellipsoid is transformed into another deformed and rotated one, $q\rightarrow g\cdot q\cdot\mbox{$^{t}g$}$. This action transforms the sphere of unit radius ($q=I$) into the ellipsoid corresponding to the positive-definite matrix $q = g\cdot\mbox{$^{t}g$}$. But any positive-definite matrix may be expressed in the form $g\cdot\mbox{$^{t}g$}$. Therefore, the general linear group acts transitively upon the family of ellipsoidal surfaces. 

With respect to the nonlinear actions considered in this article, GL(3,{\bf R}) also acts transitively upon a family of surfaces, but these are not ellipsoidal boundaries anymore. As shown in $\S3$, the nonlinear group action on open subsets $M_{\pm}$ of the Euclidean manifold is constructed by applying the nonlinear coordinate transformation $\psi_{\pm}^{-1}$ to the linear representation of GL(3,{\bf R}) in the coordinate chart ${\cal O}_{\pm}$. Hence the invariant family of  nonlinear surfaces in the Euclidean manifold is given by transforming ellipsoids in the chart space ${\cal O}_{\pm}$ back to the manifold via $\psi_{\pm}^{-1}$. If Cartesian coordinates are used for the Euclidean manifold, then each nonlinear surface in the invariant family is given by the solutions  to
\begin{equation}
1 = \mbox{$^{t}\vec{r}$}\,^{\prime} \, q^{-1}\,  \vec{r}\,^{\prime}  = \left( 1 + \frac{\Lambda}{r^3} \right)^{2/3}\, ^{t}\vec{r} \, q^{-1}\,  \vec{r} , \label{dropleteq}
\end{equation}
where  $\vec{r}\,^{\prime}$ and $\vec{r}$ denote Cartesian coordinates for ${\cal O}_{\pm}$ and the Euclidean manifold, respectively.

If $\Lambda=0$ the figures are ellipsoids. If $\Lambda$ is negative, then the ellipsoidal solutions to Eq.\,(\ref{dropleteq}) are contained entirely within the chart space ${\cal O}_{-}={\cal O}_{r\theta\phi}$. Its pre-image in $M_{-}$ is the surface of a deformed droplet excluding a spherical hole of radius $\sqrt[3]{|\Lambda|}$, see Figure 2. The half-lengths of the droplet's principal axes are $\sqrt[3]{a^3+|\Lambda|}$, $\sqrt[3]{b^3+|\Lambda|}$, and $\sqrt[3]{c^3+|\Lambda|}$. The volume of the droplet in Euclidean space, excluding the spherical bubble of volume $4\pi |\Lambda| /3$, equals the volume $4\pi a b c / 3$ of the  ellipsoid in the chart space. For $\sqrt[3]{|\Lambda|} \gg a, b, c$, the nonlinear droplet is a thin shell surrounding a spherical bubble. The width of this shell is not uniform; along a principal axis, say the $x$-axis, the width of the thin shell is approximately $(a^3/|\Lambda|)^{2/3} a/3$. In addition to exotic bubble nuclei, the negative-$\Lambda$ figures may apply to the invariant core or two-fluid nuclear model \cite{Krutov68,Gupta71}. 

If $\Lambda$ is positive, then the ellipsoidal solutions to Eq.\,(\ref{dropleteq}) may not be contained entirely within the chart space ${\cal O}_{+} = {\cal O}_{r\theta\phi}\bigcap \tilde{S}_{\sqrt[3]{\Lambda}}$. The pre-image of the portion of the ellipsoid contained in the interior of the sphere of radius ${\sqrt[3]{\Lambda}}$ is the empty set. The intersection of the ellipsoid's surface with the sphere's surface is mapped by $\psi_{+}^{-1}$ to the Euclidean space origin. The pre-image of the ellipsoidal boundary exterior to the sphere is the surface of a deformed droplet in the Euclidean manifold. In Figures 3 and 4, several positive-$\Lambda$ nonlinear surfaces are drawn for axially symmetric (a=c=1, b=2) and triaxial (a=1, b=3, c=2) shapes, respectively.  For the axially symmetric solutions, the nonlinear figures are surfaces of revolution that develop a neck as $\Lambda$ increases. At $\sqrt[3]{\Lambda}=a=c=1$, the neck is pinched down to a single point because the sphere in the chart space touches the two short principal axes of the prolate ellipsoid. For $\Lambda > 1$, the droplet's pinched neck is elongated, until, when $\sqrt[3]{\Lambda}=b=2$ (the length of the long axis of the prolate ellipsoid), the figure collapses to a point. Beyond that there is no figure in Euclidean space.

In Figure 4, the surfaces of several triaxial deformed droplets are drawn. As $\sqrt[3]{\Lambda}$ advances through the half-lengths of the principal axes $a=1$, $c=2$ and $b=3$, the pinched neck first is squeezed in the $x$-direction, then collapses to a point, and finally the figure disappears altogether. The volume of the Euclidean space droplet is the difference between the volume $4\pi a b c / 3$ of the ellipsoid in the chart space and the portion of the volume of the sphere of radius $\sqrt[3]{\Lambda}$ contained in the interior of the ellipsoid. In particular, if $\sqrt[3]{\Lambda}$ is smaller than any of the half-lengths, $a$, $b$, and $c$, then the volume of the deformed droplet equals $4\pi (a b c-\Lambda) / 3$. Some of the shapes in Figures 3 and 4 are similar to the equilibrium surfaces of a rotating charged droplet near fission \cite{WJS74}.

To determine the mechanics of a continuum fluid or finite particle system corresponding to a deformed droplet, an observable must be identified that characterizes these nonlinear geometrical shapes. Consider first a fluid contained within an ellipsoidal boundary with uniform density $\rho_{0}$. Its axes lengths and orientation are determined by its quadrupole plus monopole $Q$ tensor
\begin{equation}
Q_{ij} = \int \rho_{0}(\stackrel{\rightarrow}{r})x_{i}x_{j}d^{3}x .
\end{equation}
The inertia tensor for a rigidly rotating system is related to the $Q$ tensor by ${\cal I}_{ij} = \mbox{Tr}(Q) \delta_{ij}-Q_{ij}$. If the principal axes are aligned with the Cartesian axes, the $Q$ tensor is diagonal with entries $(M/5) a^2$, $(M/5) b^2$, and $(M/5) c^2$, where $M$ denotes the ellipsoid's mass.The orthogonal transformation that diagonalizes the $Q$ tensor is the rotation that aligns the Cartesian axes with the ellipsoid's principal axes. Hence the $Q$ tensor characterizes an ellipsoid's size and orientation.

If $\Lambda$ is negative, the surface of the deformed fluid droplet in the Euclidean manifold is mapped by $\psi_{-}$ to an ellipsoidal surface in the chart space ${\cal O}_{-}$. Suppose the density of the droplet in the Euclidean space is given by $\rho = \rho_{0}\circ \psi_{-}$ on $M_{-}$ and equals zero in the interior of the sphere of radius $\sqrt[3]{|\Lambda|}$. Then the $Q(\Lambda)$ tensor is defined by
\begin{eqnarray}
Q_{ij}(\Lambda) & = & \int \rho_{0}(\vec{r}^{\prime})x_{i}^{\prime}x_{j}^{\prime}d^{3}x^{\prime} \nonumber \\
& = & \int \rho(\vec{r}) (1 + \frac{\Lambda}{r^{3}})^{2/3} x_{i}x_{j}d^{3}x , \label{Qlambda}
\end{eqnarray}
where the first integral is over the chart domain ${\cal O}_{-}={\cal O}_{r\theta\phi}$ and the second is over the Euclidean manifold excluding the sphere of radius $\sqrt[3]{|\Lambda|}$. If the ellipsoidal surface is defined by the matrix $q$, then $Q(\Lambda) = (M/5) q$.

If $\Lambda$ is positive, the deformed droplet's surface is mapped by $\psi_{+}$ to, in general, a portion of an ellipsoidal surface in ${\cal O}_{+}$. The excluded portion is the intersection of the ellipsoid with the sphere of radius $\sqrt[3]{\Lambda}$. Suppose the density of the deformed droplet in Euclidean space is $\rho = \rho_{0}\circ \psi_{+}$. Then $Q(\Lambda)$ is defined again by Eq.\,(\ref{Qlambda}) where the first integral is over the domain ${\cal O}_{+}$ that excludes the sphere and the second is over the entire Euclidean space. But,  the correspondence between $Q(\Lambda)$ and the matrix $q$ that defines the boundary of the ellipsoid is not so simple. If $\sqrt[3]{\Lambda}$ is smaller than any of the half-lengths, $a$, $b$, and $c$, then the first integral of Eq.\,(\ref{Qlambda}) is evaluated to be
\begin{equation}
Q_{ij}(\Lambda)  = (M/5) \frac{ a b c \, q_{ij} -  \Lambda^{5/3} \delta_{ij} }{a b c -  \Lambda} .
\end{equation}

For a discrete classical system of A identical particles the $Q$ tensor simplifies to a sum over the particles
\begin{equation}
Q_{ij}(\Lambda) = \stackrel{A}{\sum_{n=1}} (1 + \frac{\Lambda}{r_{n}^{3}})^{2/3} x_{ni}x_{nj} , \label{onebody}
\end{equation}
where the mass density is replaced here by the number density. To recover the mass density expression, merely multiply Eq.\,(\ref{onebody}) by the common mass of the particles.

For a quantum system of $A$ identical particles described by a wave function $\Psi$, the expectation of the one-body operator $Q_{ij}(\Lambda)$ of Eq.\,(\ref{onebody}) equals the integral Eq.\,(\ref{Qlambda}) where the one-body density is
\begin{equation}
\rho(\vec{r}) = A \int |\Psi(\vec{r},\vec{r}_{2},\ldots \vec{r}_{A})|^{2} d^3r_{2}\cdots d^3r_{A} .
\end{equation}
The monopole and quadrupole moments of the deformed droplet are the quantum expectations of the one-body operators
\begin{eqnarray}
Q_{0}(\Lambda) & = & \sum_{n=1}^{A} (r_{n}^{3} + \Lambda )^{2/3} \nonumber \\
Q_{2m}(\Lambda) & = & \sum_{n=1}^{A} (r_{n}^{3} + \Lambda )^{2/3} Y_{2m}(\theta_n,\phi_n) , 
\end{eqnarray} 
where $Y_{2m}(\theta,\phi)$ denotes the $m$th component of the rank two spherical harmonic. If $r_n \gg \sqrt[3]{|\Lambda|}$, then the monopole, quadruple, and $Q$ tensors approximate their usual values. The Bohr-Mottelson parameters $(\beta, \gamma)$ for the deformed droplet may be defined in terms of the expectations of the quadratic and cubic scalars \cite{Kumar72,Cline88,Rose77}
\begin{eqnarray}
< [ Q_{2}(\Lambda) \times Q_{2}(\Lambda) ]^{0} > & \propto& \beta^{2} \nonumber \\
< [ Q_{2}(\Lambda) \times Q_{2}(\Lambda) \times Q_{2}(\Lambda) ]^{0} > & \propto & \beta^{3}\cos 3\gamma
\end{eqnarray}

The square of the volume of the deformed droplet is related to the expectation of the determinant of $Q_{ij}(\Lambda)$. For an ellipsoid $(\Lambda = 0)$, the square of the volume is proportional to the determinant of the $Q$ tensor
\begin{equation}
V^2 - \left( \frac{5}{A} \right)^{3} \left( \frac{4\pi}{3} \right)^{2} < \det Q > = 0 . \label{volume}
\end{equation}
If $\Lambda < 0$, then the left hand side of Eq.\,(\ref{volume}) is positive; equality is restored when $Q$ is replaced by $Q(\Lambda)$. If $\Lambda > 0$, then the left hand side of Eq.\,(\ref{volume}) is negative. The modification to restore equality is somewhat more complicated than just substituting $Q(\Lambda)$ for $Q$ and is not given here.

Hence, from the quantum expectations of the one-body operator $Q_{ij}(\Lambda)$, its quadratic and cubic scalar couplings,  and the value of the nuclear volume $V$, the matrix $q$ and the deformed droplet surface may be constructed.


\section{Nonlinear Collective Model}

The next objective is the construction of quantum collective models corresponding to the nonlinear gl(3,{\bf R}) vector fields on Euclidean space. The natural solution to the quantization problem is to determine the irreducible unitary representations of the relevant real Lie algebra of observables. An important requirement is that the Hamiltonian must be a function of the algebra observables, and, when this is true, the Lie algebra is called a spectrum generating or dynamical symmetry algebra. For example, the usual Hilbert space of square-integrable wave functions that describes a spinless nonrelativistic quantum particle is an irreducible unitary representation of the Heisenberg algebra generated by the particle position and momentum coordinate operators. For a relativistic free particle, the dynamical algebra is the Poincar\'{e} algebra spanned by the Lie algebra of the Lorentz group and the four momentum. Both are spectrum generating algebras.

The spectrum generating algebra for nonlinear collective motion surely includes the nonlinear gl(3,{\bf R}) vector fields. These observables by themselves would form a dynamical algebra if the energy were entirely kinetic. But the gl(3,{\bf R}) algebra is insufficient because the potential energy depends upon the spatial distribution of particles. Suppose the spatial distribution is defined by the nonlinear tensor $Q_{ij}(\Lambda)$ and its associated deformed droplet. Then the collective potential energy may be approximated by the sum of attractive surface tension and curvature terms plus repulsive Coulomb energies for a uniform mass and charge distribution whose boundary is the droplet's surface. These potential energies are functions of the parameter $\Lambda$ and the eigenvalues of the matrix $Q_{ij}(\Lambda)$ or, equivalently, the rotationally invariant functions of  the nonlinear $Q(\Lambda)$ tensor. This suggests constructing the algebra generated by the nonlinear gl(3,{\bf R}) vector fields and the $Q(\Lambda)$ tensor. Indeed these generators close under commutation to form the general collective motion algebra gcm(3) and, if exponentiated, the general collective motion group GCM(3). 

The general collective motion algebra is generated by two subalgebras. The first subalgebra is generated by the $Q_{ij}(\Lambda)$ tensor and is isomorphic to {\bf R}$^6$, the abelian 6 dimensional Lie algebra. The second algebra is generated by the angular and vibrational momenta, $L_{k}$, $T_{ij}(\Lambda)$, $S(\Lambda)$, and is isomorphic to gl(3,{\bf R}). The algebra gcm(3)$ \cong [{\bf R}^6]$gl(3,{\bf R}) is the span of these two Lie algebras and is a fifteen dimensional semidirect sum for which {\bf R}$^6$ is the abelian ideal. Note that two gcm(3) algebras for different $\Lambda$ are isomorphic.

A faithful matrix representation of the general collective motion semidirect sum is given by the 6$\times$6 real matrices,
\begin{eqnarray}
gcm(3) \simeq \left\{(\Xi,X)\equiv \left( \begin{array}{cc}
X & \Xi \\
0 & -^{\rm t}X \end{array}\right),\, \Xi= \mbox{$^{\rm t}\Xi$} \right\},
\end{eqnarray}
where the isomorphism is
\begin{eqnarray}
i\,Q_{ij}(\Lambda) & \mapsto & e_{i,j+3} + e_{j,i+3} \nonumber \\
i\,L_{k}/\hbar & \mapsto & \epsilon_{ijk}(e_{ij} + e_{3+i,3+j}) \nonumber  \\
i\,T_{ij}(\Lambda)/\hbar & \mapsto & e_{ij} + e_{ji} - e_{3+j,3+i} - e_{3+i,3+j} \\
& & - \frac{2}{3}\delta_{ij}(e_{11} + e_{22} + e_{33} - e_{44} - e_{55} - e_{66}) \nonumber \\
i\,S(\Lambda)/\hbar & \mapsto & e_{11} + e_{22} + e_{33} - e_{44} - e_{55} - e_{66} . \nonumber
\end{eqnarray}
e$_{km}$ denotes the 6$\times$6 matrix with 1 at the intersection of row $k$ with column $l$ and zero elsewhere.  The connected Lie group is given by exponentiation of the algebra,
\begin{eqnarray}
GCM(3) \simeq \left\{(\Delta,g)\equiv \left( \begin{array}{cc}
g & \Delta\cdot \mbox{$^{\rm t}g^{-1}$} \\
0 & ^{\rm t}g^{-1} \end{array}\right),\, \Delta = \mbox{$^{\rm t}\Delta$}, g\in GL_{+}(3,{\bf R}) 
\right\},
\end{eqnarray}
and obeys the semidirect product multiplication rule
\begin{equation}
(\Delta_{1},g_{1}) (\Delta_{2},g_{2}) = (\Delta_{1}+g_{1}\Delta_{2}\, \mbox{$^{\rm t}g_{1}$}, g_{1}g_{2}).
\end{equation}
Thus, GCM(3) is isomorphic to a semidirect product of the abelian normal 
subgroup {\bf R}$^{\rm 6}$ of 3$\times$3 real symmetric matrices under addition 
with the general linear group $GL_{+}(3,{\bf R})$.

The irreducible unitary GCM(3) representations are given by the Mackey inducing construction on associated SO(3) bundles:\newline
{\bf Theorem}.  For each nonnegative integer $C$, there exists an irreducible unitary 
representation of GCM(3) on the Hilbert space
\begin{eqnarray}
H^{C} & = & \left\{\Psi :GL_{+}(3,{\bf R})\rightarrow {\bf C}^{2C+1}\mid \right. 
\nonumber \\
& & \mbox{ } (i) \Psi(gR)=\Psi(g){\cal D}^{C}(R),\mbox{\rm for }g\in GL_{+}(3,{\bf R}), 
R\in SO(3) \nonumber \\
& & \mbox{ } (ii) \left. \int_{GL_{+}(3,{\bf R})}\parallel \Psi(g)\parallel^{2} d\nu (g) < 
\infty \right\},
\end{eqnarray}
where ${\cal D}^{C}$ denotes the $2C+1$ dimensional unitary irreducible 
representation of SO(3), and d$\nu(g)$ is the invariant measure on $GL_{+}(3,{\bf R})$.  Condition $(i)$ means that the components of the vector $\Psi(g)$ satisfy $\Psi(gR)_{K} = \sum_{K^{\prime}=-C}^{C} \Psi(g)_{K^{\prime}}{\cal D}^{C}_{K^{\prime}\, K} (R)$.
The inner product is
\begin{equation}
< \Psi | \Phi > = \int_{GL_{+}(3,{\bf R})} (\Psi(g) | \Phi(g)) d\nu (g) ,
\end{equation}
where $(v,w) = \sum_{K=-C}^{C}v_{K}^{*} w_{K}$ for $v,w \in {\bf C}^{2C+1}$.
The action of $GL_{+}(3,{\bf R})$ on $H^{C}$ is given by
\begin{equation}
(\pi(x)\Psi)(g)=\Psi(x^{-1}g),\mbox{\rm for }x,g\in GL_{+}(3,{\bf R}).
\end{equation}
The $Q$ tensor acts as a multiplication operator,
\begin{equation}
(\pi(Q_{ij})\Psi)(g) = (g\,\mbox{$^{\rm t}g$})_{ij}\Psi(g),\mbox{\rm for }g\in 
GL_{+}(3,{\bf R}) .
\end{equation}
Every unitary irreducible representation of GCM(3) is equivalent to a 
representation for some integral $C$. Two irreducible representations defined by two different integers $C$ are inequivalent.

These irreducible unitary representations model linear and nonlinear collective motion. But which models are found in nature? The answer to this question involves both kinematical and dynamical considerations. First the Hilbert space of $A$-particle states is a reducible representation of the gcm(3) algebra. A model to be kinematically allowed must be an irreducible component in the decomposition of the many-particle representation. Second, for a dynamical symmetry, the Hamiltonian must split into collective and intrinsic terms plus vanishing or small cross terms. The collective Hamiltonian acts within an irreducible representation while intrinsic terms leave invariant the perpendicular  subspace to the irreducible representation space.

The answer to the kinematic problem is that {\it every} irreducible representation of gcm(3) is a component in the decomposition of the reducible $A$-particle representation. To prove this assertion a change of variables is required for the configuration space {\bf R}$^{3A}$ from Cartesian coordinates to collective and intrinsic coordinates. Consider the linear case first ($\Lambda=0$). The general linear group acts naturally on the configuration space and, if $A>3$, almost every orbit is diffeomorphic to $GL_{+}(3,{\bf R})$. Excluding a set of measure zero from the configuration space, a  transversal may be chosen that intersects each orbit exactly once and whose points $(\bar{x}_1,\ldots,\bar{x}_A)\in {\bf R}^{3A}$ form a  $3A-9$ dimensional submanifold ${\cal N}$ for which
\begin{equation}
\sum_{n=1}^{A} \bar{x}_{na}\bar{x}_{nb} = \delta_{ab} .
\end{equation}
Then there is a vector space isomorphism
\begin{eqnarray}
{\cal L}^{2}(GL_{+}(3))  \otimes  {\cal L}^{2}({\cal N}) & \cong & {\cal L}^{2}({\bf R}^{3A}) \nonumber \\
\Psi \otimes \chi & \mapsto & f ,
\end{eqnarray}
where
\begin{eqnarray}
f(\vec{x}_1,\ldots,\vec{x}_A) & = & \Psi(g) \chi(\bar{x}_1,\ldots,\bar{x}_A) \nonumber \\
x_{ni} & = & \sum_{a=1}^{3}g_{ia}\bar{x}_{n a} ,
\end{eqnarray}
for $\Psi \in {\cal L}^{2}(GL_{+}(3))$ and $\chi \in {\cal L}^{2}({\cal N})$. This is also a Hilbert space isomorphism, i.e.,
\begin{equation}
< f_1 | f_2 > = < \Psi_1 \otimes \chi_1 | \Psi_2 \otimes \chi_2 > = < \Psi_1 | \Psi_2 > < \chi_1 | \chi_2 > . \label{innerproduct}
\end{equation}
The splitting of the inner product is a consequence of the unitarity of the representation of $GL_{+}(3,{\bf R})$ on {\bf R}$^{3A}$ and the invariance of the Haar measure on ${\cal L}^{2}(GL_{+}(3,{\bf R}))$. It follows easily that
\begin{eqnarray}
U(g) (\Psi \otimes \chi ) & = & (\pi(g)\Psi) \otimes \chi \nonumber \\
Q_{ij} (\Psi \otimes \chi ) & = & (\pi(Q_{ij})\Psi) \otimes \chi .  \label{gcmaction}
\end{eqnarray}
Thus gcm(3) acts only on ${\cal L}^{2}(GL_{+}(3,{\bf R}))$ and leaves the intrinsic wave function $\chi$ alone. Finally the reduction of this representation into irreducibles is achieved by making a polar decompostion of the general linear group into the product of a symmetric matrix times an orthogonal matrix. A basis for ${\cal L}^{2}(GL_{+}(3,{\bf R}))$ is given by the product of integrable functions on the space of symmetric matrices times the representation functions for the orthogonal group ${\cal D}^{C}_{k k'}$. Grouping together the $2C+1$ basis states for fixed $C$ and $-C\leq k' \leq +C$ provides the isomorphism with the vector valued wavefunctions of the induced representations. Since $-C\leq k \leq +C$ is not restricted, the multiplicity of each irreducible gcm(3) representation space ${\cal H}^{C}$ in ${\cal L}^{2}(GL_{+}(3,{\bf R}))$ equals $2C+1$. Because the space of intrinsic wave functions is infinite dimensional, the multiplicity of each irreducible gcm(3) representation in ${\cal L}^{2}({\bf R}^{3A})$ is countably infinite.

If $\Lambda$ is negative, a similar construction determines the GCM(3) decomposition. The principal difference from the linear case is that the separation of variables into collective $GL_{+}(3,{\bf R})$ and intrinsic transversal coordinates is made in the chart space instead of the Euclidean manifold. For $A$-particles, the coordinate system is the Cartesian product ${\cal O}_{-}^{A}$ of $A$-copies of ${\cal O}_{-}$. Removing a set of measure zero, ${\cal O}_{-}^{A}$ is diffeomorphic to the product of $GL_{+}(3)$ and the transversal submanifold ${\cal N}$. The Hilbert space isomorphism is 
\begin{equation}
{\cal L}^{2}(GL_{+}(3))  \otimes  {\cal L}^{2}({\cal N}) \cong {\cal L}^{2}({\cal O}_{-}^{A}) \cong {\cal L}^{2}(M_{-}^{A}) , \\
\end{equation}
where $M_{-}^{A}$ denotes the Cartesian product of $A$-copies of $M_{-} = {\bf R}^{3} \cap \tilde{S}_{\sqrt[3]{|\Lambda|}}$. Since the nonlinear action of  $GL_{+}(3,{\bf R})$ on ${\cal L}^{2}(M_{-}^{A})$ is unitary and the measure on ${\cal L}^{2}(GL_{+}(3))$ is invariant, the isomorphism satisfies Eq.\,(\ref{innerproduct}). The Hilbert space ${\cal L}^{2}(M_{-}^{A})$ of admissible wave functions is a subspace of ${\cal L}^{2}({\bf R}^{3A})$ consisting of the wave functions that vanish for points interior to the sphere. Since GCM(3) is represented on the tensor product space in the same way as the linear case, Eq.\,(\ref{gcmaction}), the multiplicity of each irreducible representation of gcm(3) in the decomposition of its reducible representation on ${\cal L}^{2}(M_{-}^{A})$ is again countable infinite.

If $\Lambda$ is positive, then the necessary separation of variables must be made on ${\cal O}_{+}^{A}$, the Cartesian product of $A$-copies of the chart space ${\cal O}_{+}$. But, since the chart space ${\cal O}_{+}$ excludes the sphere of radius $\sqrt[3]{\Lambda}$, the $GL_{+}(3,{\bf R})$ orbits in ${\cal O}_{+}^{A}$ are not globally diffeomorphic to the Lie group $GL_{+}(3,{\bf R})$. If $i$ denotes the natural injection of ${\cal O}_{+}^{A}$ into {\bf R}$^{3A}$, then the Hilbert space of states is ${\cal L}^{2}(Q)$, where the manifold $Q = i^{-1} ( GL_{+}(3)\times N)$. Although the regular representation of $GL_{+}(3)$ is not defined anymore because of the sphere's obstruction, the Lie algebra $gl(3,{\bf R})$ representation, dependent upon the local group action, is well-defined on ${\cal L}^{2}(Q)$. Since the injection $i$ commutes with the derived Lie algebra representation, its decomposion into irreducibles is identical to the previous cases.

For a spectrum generating algebra the collective Hamiltonian must be a function of the gcm(3) Lie algebra elements. A phenomenological gcm(3) Hamiltonian is provided by a linear combination of rotational scalars formed from the generators. For a  tractable model the scalars may be restricted to polynomials of low degree. Up to quadratic, time-reversal symmetric terms, the kinetic energy  is a sum of rotational, quadrupole vibrational, and breathing oscillation terms constructed from the general linear group generators
\begin{equation}
T = \frac{L^2}{2 {\cal I}} + \frac{ T(\Lambda)\cdot T(\Lambda) }{2B} + \frac{S(\Lambda)^2}{2C} ,
\end{equation}
where ${\cal I}$, $B$ and $C$ are adjustable parameters. A more realistic model would include higher degree polynomials including terms like the cubic scalar $[T(\Lambda) \times Q^{(2)}(\Lambda) \times T(\Lambda)]^{0}$.

The potential energy is a scalar function of the quadrupole and monopole components of the $Q(\Lambda)$ tensor. The quadrupole scalars are functions of the quadratic $[Q^{(2)}(\Lambda) \times Q^{(2)}(\Lambda) ]^{0}\propto \beta^2$ and cubic $[ Q^{(2)}(\Lambda) \times Q^{(2)}(\Lambda) \times Q^{(2)}(\Lambda) ]^{0} \propto \beta^3 \cos 3\gamma$ scalars. Hence, the quadrupole part of the potential may be expressed as a function $V(\beta,\gamma)$. The monopole component of the potential should have a $Q^{0}(\Lambda)$ term. For incompressible motion there may be a $\det Q(\Lambda)$ term too.

\section{Conclusion}
The results of this paper may be amplified, developed and extended in several directions. First, in addition to the $GL_{+}(3,{\bf R})$ actions of this article, other nonlinear group actions on Euclidean space may be discovered and applied to the description of geometrical collective motion in many-particle systems. The nonlinear $GL_{+}(3,{\bf R})$ collective motion theory developed here is an example of the constructive procedure for any nonlinear theory of geometrical modes. Although the requirement that the set of collective vector fields form a Lie algebra that contains the angular momentum algebra $so(3)$ may not be relaxed, a finite dimensional algebra is not essential provided the theory is still tractable, e.g., the collective vector fields might be a Virasoro or Kac-Moody algebra.

Second the adjustable parameters of the collective potential energy may be determined from more basic considerations than just phenomenological fitting to energy levels and transition rates. For example, the potential energy may be defined by attractive surface and curvature energies plus Coulomb repulsion for a uniform positively charged fluid contained within the surface of the deformed droplets of \S4. More fundamentally, the collective energy may be derived from a microscopic Hamiltonian $H$ after an intrinsic wave function $\chi$ is chosen. One way to give such wave functions is by projection from a Hartree-Fock state $\psi_{HF}$,
\begin{equation}
\chi(\vec{x}_{1},\ldots, \vec{x}_{A}) = \int_{GL_{+}(3,{\bf R})} d\mu(g)\ \psi_{HF}(g^{-1}\cdot \vec{x}_{1},\ldots, g^{-1}\cdot \vec{x}_{A}) .
\end{equation}
Since $\mu$ is the invariant measure on the general linear group, $\chi$ is an intrinsic wave function, i.e., $\chi(g\cdot \vec{x}_{1},\ldots, g\cdot \vec{x}_{A}) = \chi(\vec{x}_{1},\ldots, \vec{x}_{A})$ for every $g\in GL_{+}(3,{\bf R})$ and $\chi$ is well-defined on ${\cal N}$. The matrix elements of the collective Hamiltonian $H_{c}$ are defined by
\begin{equation}
< \Phi | H_{c} | \Psi > = < \Phi \otimes \chi | H | \Psi \otimes \chi > ,
\end{equation}
where $\Phi, \Psi \in {\cal L}^{2}(GL_{+}(3,{\bf R}))$.

Finally the nonlinear gcm(3) algebra may be extended to the symplectic Lie algebra sp(3,{\bf R}). This algebra consists of the quadratic functions of the position and momentum operators or, equivalently, the quadratics in the oscillator phonons. For the nonlinear geometries this algebra should be defined using the chart space coordinates of ${\cal O}_{\pm}$. The position and derivative operators in the chart coordinates $(x_1^{\prime},x_2^{\prime},x_3^{\prime})$ are related to the corresponding operators on the Euclidean manifold by
\begin{eqnarray}
x_{i}^{\prime}  & = & (1+\frac{\Lambda}{r^3})^{1/3} x_{i} \nonumber \\ 
\frac{\partial}{\partial x_{i}^{\prime}} & = & (1+\frac{\Lambda}{r^3})^{-1/3} \left[ \frac{\partial}{\partial x_{i}} -\frac{\Lambda x_{i}}{r^{4}} \frac{\partial}{\partial r} \right] .
\end{eqnarray}
Defining the nonlinear phonons by $a_{i}^{\prime} = (x_{i}^{\prime} + \partial / \partial x_{i}^{\prime} )/\sqrt{2}$, the complexification of the real symplectic Lie algebra is 
\begin{equation}
sp(3,{\bf R}) = \mbox{span} \left\{ a_{i}^{\prime}a_{j}^{\prime} , (a_{i}^{\prime})^{\dagger}(a_{j}^{\prime})^{\dagger} , [ (a_{i}^{\prime})^{\dagger}a_{j}^{\prime} + (a_{j}^{\prime})^{\dagger}a_{i}^{\prime}]/2  \right\} . 
\end{equation}
Its maximal compact subalgebra is the nonlinear Elliott $u(3)$ Lie algebra
\begin{equation}
u(3) = \mbox{span} \left\{ [ (a_{i}^{\prime})^{\dagger}a_{j}^{\prime} + (a_{j}^{\prime})^{\dagger}a_{i}^{\prime}]/2  \right\} .
\end{equation}
The irreducible representatations of the nonlinear symplectic $sp(3,{\bf R})$ and the nonlinear Elliott $su(3)$ algebra for $A$-particles are given by highest weight arguments similar to the usual linear symplectic and Elliott theories, albeit modified to the chart space ${\cal L}^{2}({\cal O}_{\pm}^{A})$. These adapted symplectic and Elliott models are the natural frameworks for microscopic calculations of nonlinear collective motion.

\newpage
{\bf List of Figures}
\begin{enumerate}
\item Figure 1. Coordinate charts for three-dimensional Euclidean space.
\item Figure 2. For $\Lambda = -1$ and $a=b=1, c=2$, the droplet is a surface of revolution that contains a spherical cavity. 
\item Figure 3. When $a=c=1, b=2$ and $\Lambda$ is positive, the droplet is a surface of revolution with a neck.
\item Figure 4. Triaxial droplets for various positive $\Lambda$ and $a=1, b=3, c=2$. Sections through the principal $XZ$, $YZ$, and $XY$ planes are drawn for each droplet at the nine, twelve, and three o'clock positions, respectively. 
\end{enumerate}
\end{document}